# 3D Primitives Gpgpu Generation for Volume Visualization in 3D Graphics Systems

## Anas M. Al-Oraiqat[1] and Sergii A. Zori[2]

*[1]Department of Cyber Security, Onaizah Colleges, College of Engineering & Information Technology، Onaizah, Kingdom of Saudi Arabia, P.O. Box 5371, and [2]Department of Computer Sciences and Technologies, SHEE «Donetsk National Technical University», Donetska Oblast, P.O.Box 85300, Ukraine*

anas_oraiqat@hotmail.com

*Abstract*. This article discusses the study of 3D graphic volume primitive computer system generation (3D segments) based on General Purpose Graphics Processing Unit (GPGPU) technology for 3D volume visualization systems. It is based on the general method of Volume 3D primitive generation and an algorithm for the voxelization of 3D lines, previously proposed and studied by the authors. We considered the Compute Unified Device Architect (CUDA) implementation of a parametric method for generating 3D line segments and characteristics of generation on modern Graphics Processing Units. Experiments on the test bench showed the relative inefficiency of generating a single 3D line segment and the efficiency of generating both fixed and arbitrary length of 3D segments on a Graphics Processing Unit (GPU). Experimental studies have proven the effectiveness and the quality of produced solutions by our method, when compared to existing state-of-the-art approaches.

## 1. Introduction

One of the most promising but little addressed by the development of 3Dcomputer graphics systems and visualization is the transition to a "true" 3D volume visualization [1-7]. It is required to have the appropriate equipment construction for 3D volume visualization (3D displays) and the fundamental new development methods/algorithms for generating 3D volume (voxel) primitives for them [1, 2, 8-14].

The authors in [1, 2, 15, 16] proposed algorithmic approaches and methods that aim at organizing the voxel representation generation of graphic 3D primitives for volumetric context display devices. The operations uniformity in the proposed approaches and algorithms for 3D visualization made it possible to achieve their relatively simple implementation and good acceleration on high-performance Graphics Processing Unit (GPU) systems. GPU Systems are widely used today in parallel implementations of Single-Program Multiple-Data (SPMD) tasks, both for general purpose and non-classical (not performed on a standard 3D graphic pipeline) computer graphics tasks [1, 2, 6, 7, 13, 14].

This article is the authors' continuation research on creating a framework for 3D

volume visualization based on General Purpose Graphics Processing Unit (GPGPU), to get a hardware support for accelerating the volumetric 3D images synthesis. It also discusses implementing and generating characteristics for basic 3D volume graphics primitive of the 3D straight-line segment, on modern GPUs.

The reminder of this paper is organized as follows: Section 2 discusses the related work. Section 3 presents the general algorithm of generating a 3D line segments. In section 4 and section 5 respectively, we show the Compute Unified Device Architect (CUDA) implementation of 3D line segments generation, and we provide the experimental results. Finally, Section 6 is devoted to the conclusion and the future work.

## 2. Related Works

The main idea of voxel graphics is to represent the scene as a 3D voxels array, the sides of which are aligned along the coordinate axes. The main task of the volumetric image synthesis is to search for voxels in the scene voxel volume. That accurately approximates the scene objects 3D surfaces. This approach is called Direct Volume Rendering [1, 2, 8-12].

There are two classes of volume visualization methods in primitives' terms. For the methods based on the objects representation by surfaces [8], an intermediate model is first created with the object surface highlighted. Next, the surfaces are rendered. Other methods based on voxel representation of volumes create 3D object images directly from the volume data of 3D graphics primitives [1, 2, 8-12]. Both of these methods have their merits, and one of them should be used for a particular application and it depends on the visualization objectives.

Unfortunately, the lack of commercial solutions regarding 3D volume displays necessitates the output of volumetric images to traditional display devices. That is connected with the need for additional 2D rasterization of the resulting voxel model and leveling the "true 3D volume visualization" advantages [2, 8, 11-14]. Note that, in the literature, methods of voxel models creating real objects are not considered in details. Such models are usually obtained automatically as a result of 3D scanning, etc. For example, using 3D scanners, while methods and algorithms for generating typical 3D volume graphics primitives practically are not considered [2, 8, 12].

In [10], the authors considered the representation of the general volume visualization theory, the basic 3D Graphics Primitive used for Volume Visualization, and a searching for a general algorithm for voxels belonging to a graphic primitive. However, algorithms for their generation and the comparison with other methods are not available.

In [11], the authors have focused on Voxel Primitive Modeling and Simulating (VPMS) problems, the general structure of the VPMS system was proposed. The significance of effective voxel 3D Graphics Primitive generation within it are determined. However, specific methods and algorithms for their generation are not given.

In [12], a general algorithm is proposed for searching voxels belonging to a volumetric graphic primitive. The authors performed a spatial errors analysis of 3D volume generation, but specific implementations of basic 3D Graphics Primitive algorithms are not considered.

In [13], the voxels rendering framework was designed and implemented in 3D, with a focus on sparse 3D graphics, and using 3D geometry shaders on GPU. The proposed approach reduces the CPU-to-GPU bandwidth costs of updating the volumetric grid data. That makes it more suitable for dynamic voxel-based environment rendering GPU rasterization as opposed to ray-casting oriented approaches. However, the need to get a 3D image on a 2D display means the vixelization of the voxel model, and does not apply to a real 3D volume visualization.

In [14], parallel GPU based visualization of voxelized surface Data was taken into consideration, but the task and algorithms for generating volumetric 3D Graphics Primitive have not been considered.

In [1, 2, 15, 16], the authors considered 3D graphics primitive generating problems, particularly the 3D straight segments studied in this work. The authors provided methods and algorithms to solve them. The volume decomposition task of a line segment is formulated as the determining task voxels set, each of which (except the initial and final) has two and only two adjacent voxels, with each center lying at the minimum distance to the 3D line segment.

In [1, 16], a method to get such a set is proposed. The essence of which, is that at a certain generation step there is a specific voxel sequence, with the requirement to determine the decomposition of the next voxel. To do this, it considered seven neighboring voxels-applicants in the direction that is determined by the straight segment-directing vector. The next voxel in the sequence is defined as a voxel-applicant with a minimum distance to a given straight line. On this basis, the authors proposed several 3D Graphics Primitive generation algorithms for a straight segment, where they evaluated and compared their efficiency. It is shown that the "Parametric Method" possesses the best characteristics and potential parallelization in GPGPU implementation that is investigated in this paper. A summary of the above discussed works is presented in Table 1.

From the above discussion, we conclude that researchers have dealt with various volume visualization issues, like voxelization of geometric scenes, voxelized surface data visualization, etc. Also, we can notice, from Table 1, that 3D volume displays is the most focused data type in most approaches, except for [13] which concentrated their efforts on 3D volume on 2D displays. As for the voxelization constraints, some authors used the volume scene information, like in [8- 12], while others have used different constraints, such as volumetric grid data [12] and boundary volume information [14]. We also noticed, from Table 1, that existing solutions have been implemented with different methods depending on the treated problem and the focused constraints. Examples of those methods include Ray-casting oriented rendering on GPU [13], primitive voxelization [10, 12], etc.

However, a common limitation between the above-discussed works is that the generation techniques for typical 3D volume primitives have not been considered. In addition, the reduced performance and the quality of produced solutions are common issues in existing works. In this paper, we give a solution to the above challenges by applying a parametric method. This later will ensure the parallelization in GPGPU implementation, which will positively affect the process of 3D volume generation.

## 3. 3D Volume Primitives and 3D Line Segment Generation

Volumetric spatial visualization systems are usually built considering the general concept of discrete voxel scene generating 3D volumetric images, which is described by the authors in [1, 2, 7]. The generalized structure of such system is shown in Fig. 1.

**Table 1. Approaches comparison.**

| Approach | Goal | Data type | Constraints | Adopted techniques |
|---|---|---|---|---|
| [8-9] | Volume sampled voxelization of geometric scenes | 3D volume displays | Volume scene information | Method of voxelization |
| [10] | General volume visualization | 3D volume displays | Volume scene information | Algorithm for primitive's voxelization |
| [11] | Voxel Primitive Modeling and Simulating | 3D volume displays | Volume scene information | VPMS system structure |
| [12] | Spatial errors analysis of 3D volume generation | 3D volume displays | Volume scene information | Algorithm for primitive's voxels searching |
| [13] | Voxels rendering using 3D geometry shaders on GPU | 3D volume on 2D displays | Volumetric grid data | Ray-casting oriented rendering on GPU |
| [14] | Parallel GPU based voxelized surface data visualization | 3D volume displays | Boundary volume information | GPU based voxelized data visualization |
| **Our approach** | Fast parallel and GPU based voxelization | 3D volume displays | Volume scene information | Parametric Method |

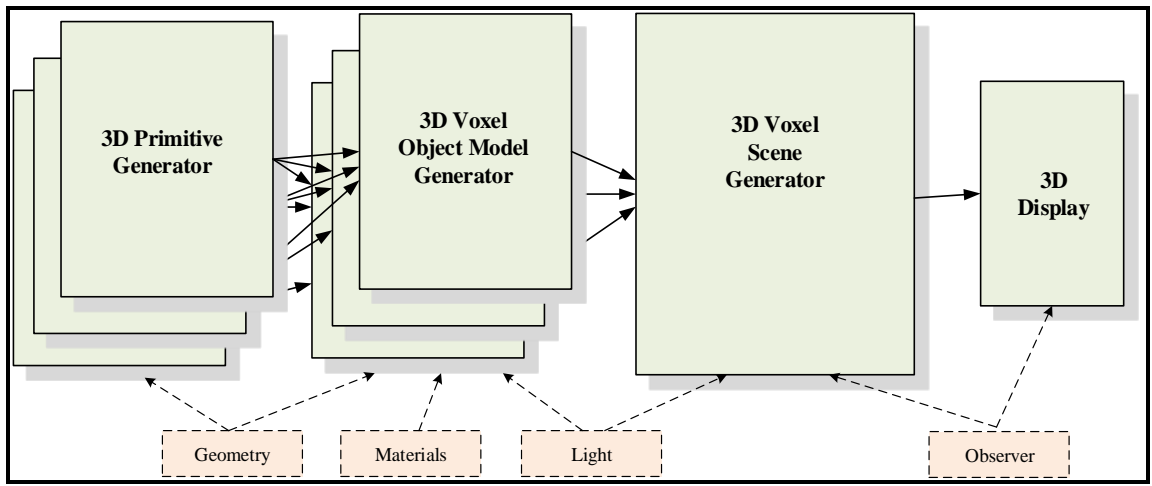

**Fig. 1. Computer architecture of 3D volume visualization.**

One of the main system components is the basic 3D volume primitive generator. Its resulting outputs are considered, later, in the "assembly" of 3D voxel models of objects and scene. In this regard, the synthesis accelerations' main directions in the systems that belongs to this class are associated with:

- The effective generation of basic 3D volume primitive's voxel representations.

- The acceleration of their generation on the proposed algorithmic base, using the mapping on the architecture of a parallel computing system.

In [1, 2, 15, 16], the authors proposed algorithmic approaches and methods of the voxel generation's representation for basic graphic 3D primitives. These methods intended for volumetric display devices of context and based on the developed generalized method for generating an arbitrary volume 3D primitive [1, 2].

The method essence and the generalized method of generating an arbitrary volume 3D primitive as shown in Fig. 2 is described in details in [1]. $\Gamma(x,y,z)$ is an arbitrary 3D volume graphics primitive, $P(x,y,z)$ is a surface which contains $\Gamma$, $\Phi(x,y,z)$ is a closed curve which bounds P within $\Gamma$.

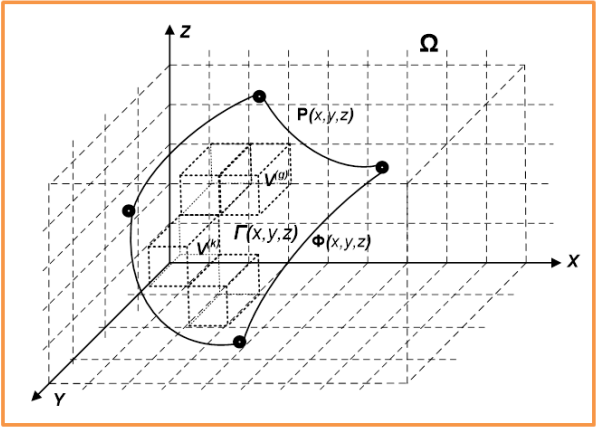

**Fig. 2. Illustration of generating an arbitrary 3D volume primitive $\Gamma$ [1].**

On the basis of the generalized method for generating an arbitrary 3D volume primitive, the authors also developed and theoretically investigated algorithms for generating an arbitrary 3D line segment as a basic 3D-graphic volume primitive of [1, 16]. The experiments have shown that the "Parametric Method" algorithm has the best performance characteristics in terms of speed [1]. The essence of the "Parametric method" is shown below and illustrated in Fig. 3.

An arbitrary point **G** of the line segment can be determined parametrically as:

$$\mathbf{G} = \mathbf{E}\,t + \mathbf{S}(1-t) = \mathbf{S} + (\mathbf{E}-\mathbf{S})\cdot t,\ \ 0 \le t \le 1$$

where S and E – Start and End points of line segment and $t$ is a parameter.

If $t$ is computed as

$$t_k = \frac{1}{\|\mathbf{E}-\mathbf{S}\|}k,\ \ k = 0, 1, 2, ..., \text{int}(\|\mathbf{E}-\mathbf{S}\|)$$

then the points $\mathbf{G}_k$ will be consistently placed into all voxels that may belong to the 3D decomposition of the $\mathbf{SE}$ segment [1].

$$\mathbf{G}_k = \mathbf{S} + \frac{(\mathbf{E}-\mathbf{S})}{\|\mathbf{E}-\mathbf{S}\|}k,\ \ k = 0,1,2,...,\text{int}(\|\mathbf{E}-\mathbf{S}\|)$$

(1)

The parameter $t$ in (1) with considering the specificity of N ( $N = \text{int}(\|\mathbf{E}-\mathbf{S}\|)$ should be set as [1]:

$$t_k = \frac{1}{\|\mathbf{E}-\mathbf{S}\|} \cdot \frac{\|\mathbf{E}-\mathbf{S}\|}{N}k = \frac{1}{N}k,$$
$$k = 0,1,2,...,N$$

(2)

and

$$\mathbf{G}_k = \mathbf{S} + \mathbf{W}\cdot k,\ \ k = 0, 1, 2, ..., N,\ \ \mathbf{W} = \frac{\mathbf{E}-\mathbf{S}}{N}$$

(3)

In case of selection of $t_k$, $\mathbf{G}_k$ points, it will consistently run through all unknown voxels of the $\mathbf{SE}$ line segment 3D raster decomposition.

**"Parametric Method" Algorithm**

**Input data:** $P(x, y, z)$, $\Phi(x, y, z)$

**Output data: Array of voxels** $V^{(k)}$

**begin**

**Initialize point S and end point E;**

**Defining** $N$ ;

**Computation of W using (4) ;**

**for k:=1 to N do**

$$V_{SE}^{(k)} \supseteq \mathbf{S} + \mathbf{W}\cdot k$$

**end_for**

**end**

**Fig. 3. General algorithm "Parametric Method" of generating a 3D line segment [1].**

The proposed parametric method has a large potential for parallelization because the search for a certain voxel of decomposition with such approach does not depend on the search for neighboring voxels.

## 4. Generation of 3D Line Segments on the GPU

Consider the organization of GPU computations for generating a 3D line segment based on the developed parametric method [1]. Since the search for some voxel decomposition does not depend on the search for neighboring voxels, the following organization of the computational process GPGPU is proposed, as shown in Fig. 4(a).

The implementation method consists of the following steps:

1) CPU preprocessing (computation of the length of segment line, the voxels number in the segment and the guiding vector).

2) CPU allocation of memory for storing initial and final voxels.

3) GPU global memory allocation for calculating voxels on a video card.

4) Grid configuration of blocks and GPU threads.

5) CUDA parallel core function call (its parameters are auxiliary data computed at the first stage).

6) Synchronization of blocks and threads: the kernel function is asynchronous, so it is needed to make sure that the threads of all the blocks have completed the computations.

7) Copying the computed data from the GPU memory to the CPU memory.

8) Release of all previously allocated memory (CPU, GPU).

The parallel core function is launched in N blocks, each of which computes one voxel's vertex coordinate. To transfer data between the CPU and GPU, the voxels coordinates are saved in the global memory of the video card during the computations. Once these latter are achieved, the data is copied to the host. Next, we consider both, the capabilities and generating characteristics as a single 3D line segment on the GPU (see Fig. 4a), and the multitude simultaneous generation of arbitrary volume line segments (see Fig. 4b).

For GPU generation of multiple 3D segments, some preliminary computations are also required - computing the length of each straight line $Ni$ and the maximum length of the straight line $Nmax$. In addition, $Ui$ is computed - the direction vector of the $i^{th}$ segment. After memory is allocated, the CUDA grid is configured for all the generated lines.

## 5. Experiments

With the parallel generation of one segment on the GPU by the above display (Fig. 4a), the blocks number in the X dimension is responsible for the voxels number in the straight segment. Nevertheless, generating multiple segments (Fig. 4b) in one core function makes this measurement contains the maximum possible number of voxels in the generated segments $Nmax$.

In the core function, before computing the current voxel coordinates, first, it is determined whether the voxel is one of the generated segment voxels (its number is less than the segment's length), or whether it is a redundant. Then, $W$ and the voxel coordinates are computed for this voxel. In the voxel decomposition case of multitude segments, the characteristics were experimentally evaluated when generating $NP$ line segments at the same time. However, all segments' generation on the GPU is performed in parallel, in one core function call.

Possible options for the grid configuration for a large threads number were considered:

- $N_P$ blocks on 1 stream.

- $N_P/N_{TH}$ blocks by $N_{TH}$ flows (by dimension y).

If $N_P$ is greater than the maximum possible block size and this is necessary for solving real problems of large dimension, the option of one block $N_P$ stream is excluded. Then $N_{TH}$, the threads number in the block, is limited by the capabilities of a particular GPU [17]. The $N_{TH}$ choice must be made so that the total streams number across all measurements is at least 64 (2 warps). It is also necessary to take into account the shared amount of memory used by the core function.

After generating $N_P$ straight lines, the size of this data increases $N_P$ times. To transfer them to the core function, it is necessary to allocate memory on the GPU. It is desirable to use the constant video card memory, as it is faster and more cacheable than the global one.

When conducting experiments, the following parameters were varied:

1) A voxel decomposition of one line segment of a given length from 1000 to 1000000 voxels is performed.

2) A voxel decomposition of multiple 1024 fixed-length line segments from 20 to 200000 voxels is performed.

3) A voxel decomposition of 1024 arbitrary straight-line different lengths segments with a total one Giga-Voxels length is performed. If all segments have different lengths, then the streams number (blocks) determining the voxels number in each of the generated segments is equal to the maximum length of all segments. And in the computations, diverging streams inevitably arise. That negatively affects the overall speed. In the worst case, the result was accepted for an overall assessment.

4) Time intervals were measured using the NVIDIA CUDA Visual Profiler.

5) For the tested configuration, the solution was implemented on a Windows 7 Ultimate x64/Intel® Core™2 Quad Q9550 2.83 GHz and 4 GB DDR3. Four GPU types are considered [12]:

- NVidia GeForce GTX 260 (192 CUDA Core).

- NVidia GeForce GTX 1050 (640 CUDA Core).

- NVidia GeForce GTX 760 (1152 CUDA Core).

- NVidia GeForce GTX 1070i (2432 CUDA Core).

Note that the GeForce GTX 260 is used also in the previous experiments, described in [1], while the GeForce GTX 760 and GTX 10x [17] are working in the backward compatible PCI-e 2.0 x16 interface due mode, to be applied to the test bench motherboard features. This, somehow, limited their overall maximum performance in the system. Comparative implementations data, using test bench resources for the voxelization of single-line segments, are summarized in Table 2 and are shown in Fig. 5 and 6.

The times of voxel decomposition results of a multi 1024 3D line segments are shown in Table 3 and in Fig. 7 and 8. Table 4 and Fig. 9 show the Parametric Method Generation performance with and without the GPUs test.

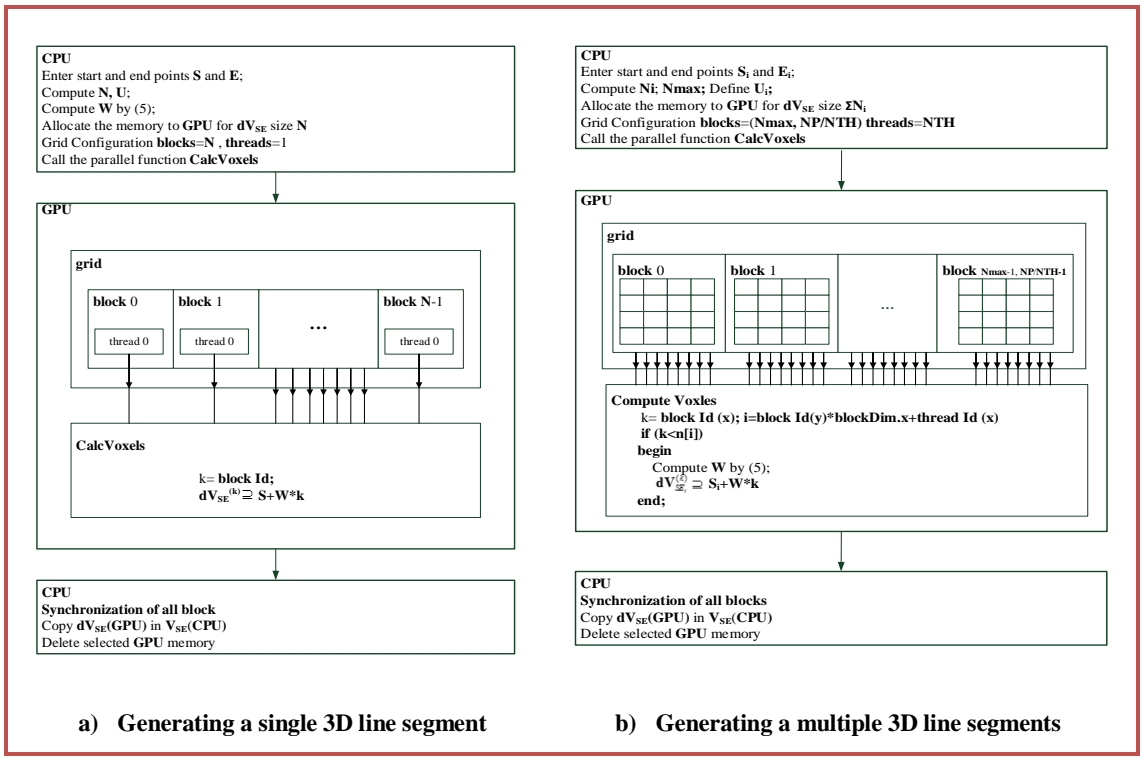

a) **Generating a single 3D line segment**     b) **Generating a multiple 3D line segments**

**Fig. 4. Hybrid (CPU + GPU) implementation of "Parametric Method"** [1]**.**

**Table 2. Experimental results for the generation of a single segment.**

| Straight length, Vox. | Implementation time, (ms) | | | | |
|---|---|---|---|---|---|
| | CPU Q9550 | GPU GTX 260 | GPU GTX 1050 | GPU GTX 760 | GPU GTX 1070i |
| 1000 | 0.05 | 1.84 | 1.57 | 0.91 | 0.45 |
| 2000 | 0.08 | 1.86 | 1.58 | 0.92 | 0.46 |
| 5000 | 0.16 | 1.99 | 1.69 | 0.99 | 0.49 |
| 10000 | 0.29 | 2.04 | 1.74 | 1.01 | 0.51 |
| 20000 | 0.59 | 2.07 | 1.77 | 1.02 | 0.52 |
| 50000 | 1.43 | 2.53 | 2.16 | 1.27 | 0.67 |
| 100000 | 2.73 | 3.49 | 2.98 | 1.75 | 0.92 |
| 200000 | 5.24 | 5.49 | 4.86 | 3.05 | 1.52 |
| 500000 | 8.74 | 7.89 | 7.01 | 4.51 | 2.22 |
| 1000000 | 16.3 | 12.79 | 11.42 | 7.31 | 3.65428571 |

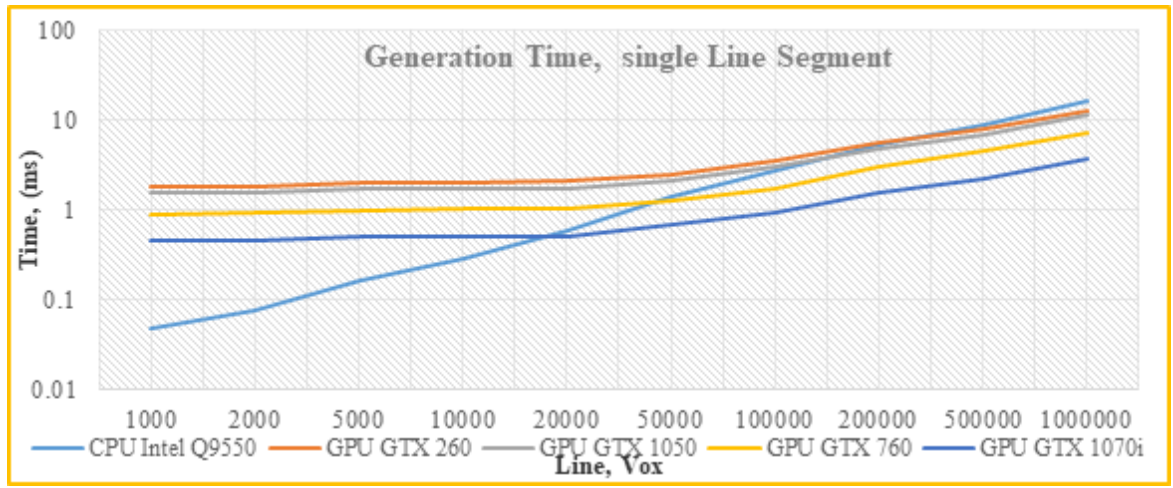

**Fig. 5. CPU and GPU realization of the "Parametric Method", a single 3D line segment.**

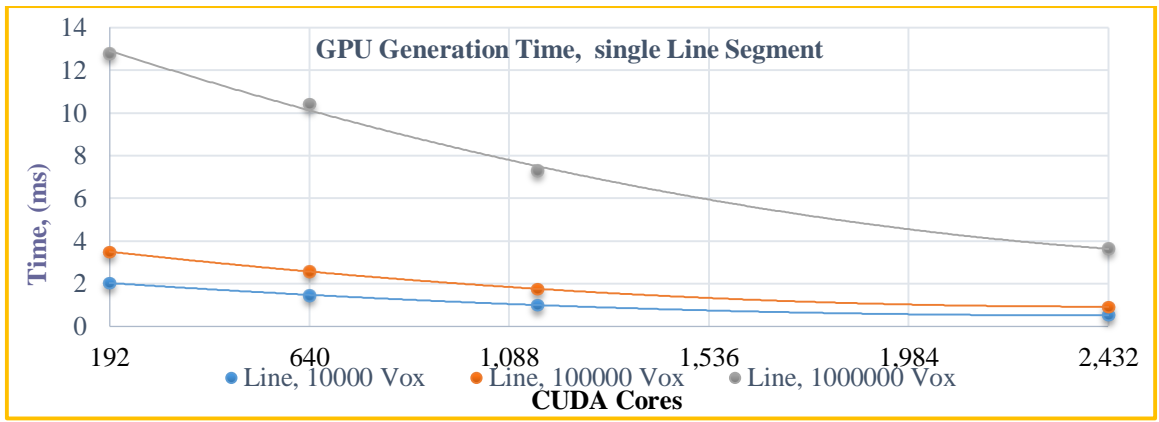

**Fig. 6. GPU implementation of the "Parametric Method", a single 3D line segment.**

**Table 3. Time of voxel decomposition on the CPU and GPU for 1024 straight line segments of the same length.**

| Straight length, Vox | Implementation time, (ms) | | | | |
|---|---|---|---|---|---|
| | CPU Q9550 | GPU GTX 260 | GPU GTX 1050 | GPU GTX 760 | GPU GTX 1070i |
| 20 | 3.2 | 3.42 | 1.46 | 0.84 | 0.42 |
| 50 | 4.11 | 3.8 | 1.62 | 0.94 | 0.46 |
| 100 | 7.09 | 5.97 | 2.54 | 1.48 | 0.75 |
| 1000 | 26.23 | 15.82 | 4.74 | 2.63 | 1.44 |
| 2000 | 51.38 | 33.03 | 14.12 | 8.14 | 4.13 |
| 5000 | 124.44 | 82.35 | 35.19 | 20.59 | 10.91 |
| 10000 | 240.03 | 163.87 | 70.03 | 40.97 | 21.70 |
| 20000 | 474.58 | 320.07 | 141.62 | 88.91 | 43.85 |
| 50000 | 1417.4 | 783.18 | 348.08 | 223.77 | 105.84 |
| 100000 | 2640.33 | 1527.20 | 681.79 | 436.34 | 218.17 |
| 200000 | 5315.296 | 2993.31 | 1336.30 | 855.23 | 427.62 |

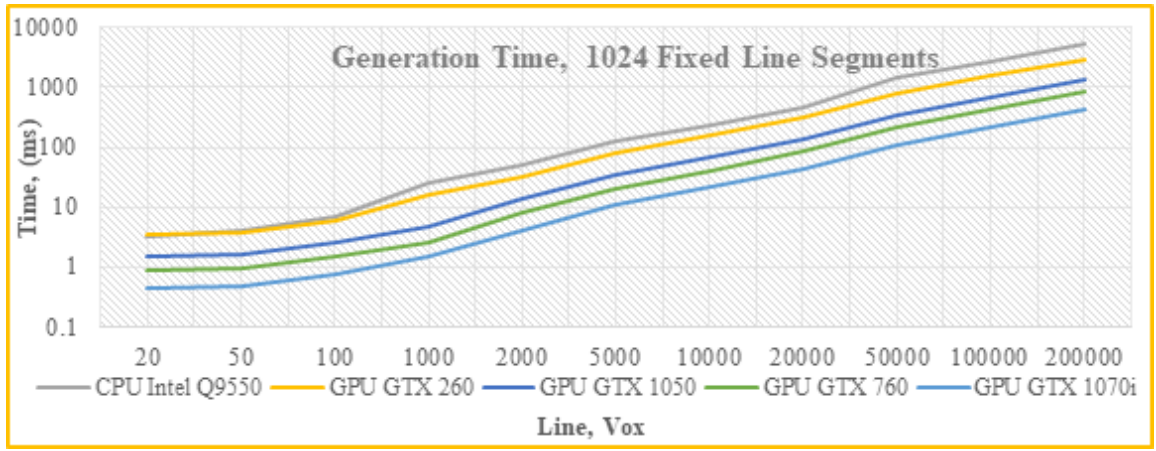

**Fig. 7. CPU and GPU implementation of the "Parametric Method", 1024 3D fixed-length line segments.**

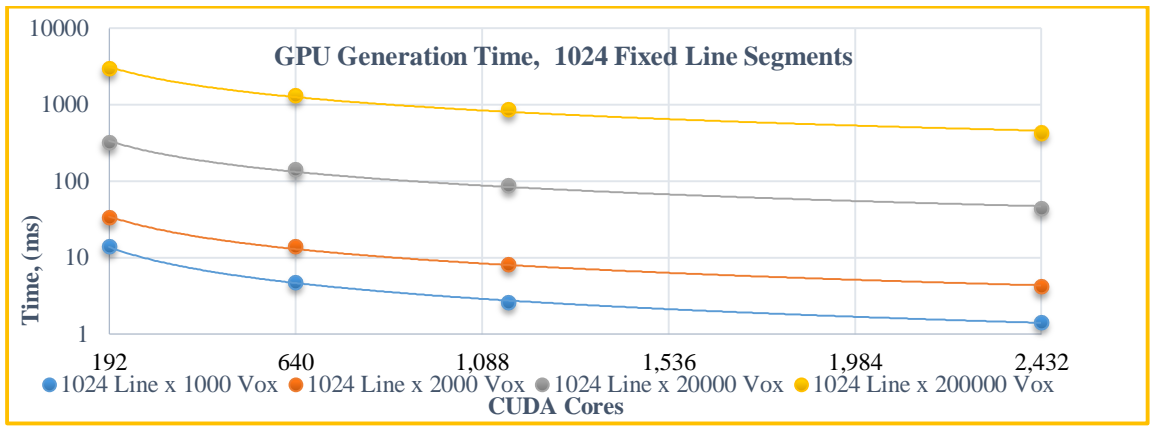

**Fig. 8. GPU implementation of the "Parametric Method", 1024 3D line segments of fixed length.**

**Table 4. Time and performance of the voxel expansion on the CPU and GPU for a set of segments of arbitrary length.**

| Total all straight lines length | Implementation time, ms | | | | |
|---|---|---|---|---|---|
| | CPU Q9550 | GPU GTX 260 | GPU GTX 1050 | GPU GTX 760 | GPU GTX 1070i |
| 1 Giga-Voxels | 119729.2 | 29932.29 | 13000.30 | 8051.23 | 4071.01 |
| | Performance (MVs) | | | | |
| | 8.4 | 33.4 | 77.2 | 124,2 | 245.7 |

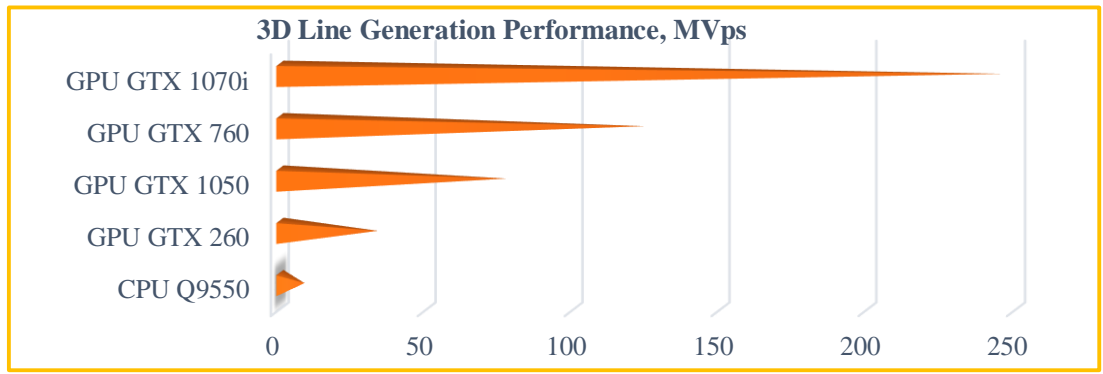

**Fig. 9. Generation performance of the "Parametric method", arbitrary 3D line segments with a total one Giga-Voxel length.**

The comparative analysis of the data obtained via the conducted experiments allowed us to draw the following conclusions:

1. The parametric method implementation of generating a single 3D line segment has the following characteristics:

   - Computing resources of the test system video cards are used by 25-30%.

   - Around 70% of the computation time it spent to copy the computed data to the host. It takes place because the computations practically do not contain branches and contain a reduced number of simple computational operations.

2. Parallel GPU implementation of the single 3D line segment generation, practically does not give a gain in time, compared with the CPU implementation for small-length segments (up to 100000 voxels). This is because most of the time is not occupied by the computations themselves, but by copying data from memory between the host and the video card and synchronizing the streams. When generating large segments, the situation improves, but the acceleration per unit of the GPU computational core is still small.

3. Multiple sequential generations of single segments require multiple calls to the function-core and the corresponding data load/unload number of operations from/to the video card memory, which also prevents the efficient use of GPU resources. Therefore, for an efficient use of GPU resources, it is necessary to limit the calls number to the function-kernel and at the same time increase the load on the GPU through simultaneous GPU voxel decomposition of a multitude segments.

4. The parametric implementation method for generating multi 3D line segments, allows performing a full computational GPU load (up to 100%). However, it still takes a significant amount of time to copy the computed data to the host that limits the potential acceleration. When generating a multi arbitrary-length 3D segments. The acceleration is also hampered by the presence of branching and branching streams in the blocks, while simultaneously generating different 3D segments.

5. The total performance shown on the test bench in the CUDA architecture, when generating 1 GVox line segments, is 245 MVps which exceeded the CPU performance by 30 times.

## 6. Conclusions

In this paper, the problem of the 3D line segments voxelization and its characteristics on modern GPUs was studied. We proposed an algorithm for the voxelization of 3D lines and its CUDA implementation on GPU. Our aim was to confirm the effectiveness of the developed method for generating 3D line segments and the possibility of the real-time GPGPU generation of 3D segments with a high total length. The obtained results have shown that the total performance of the CUDA implementation exceeded the CPU performance about 30 times.

Future work will be devoted to the improvement of the proposed algorithm, in order to increase the efficiency of its parallel implementation in GPGPU.

**Conflict of Interest:** The authors declare that they have no conflicts of interest.